%% file: main.tex
\documentclass[submission,copyright,creativecommons]{eptcs}
\providecommand{\event}{SYNT 2016} 

\usepackage{cite}
\usepackage{algorithm}
\usepackage[noend]{algpseudocode}
\usepackage{amsfonts}
\usepackage{amsmath}
\usepackage{amssymb}
\usepackage{array}
\usepackage{appendix}
\usepackage{colonequals}
\usepackage{xcolor}
\usepackage{courier}
\usepackage{pifont}
\usepackage{defs}
\usepackage{enumitem}
\usepackage{graphicx}
\usepackage{latexsym}
\usepackage{listings}
\usepackage{mathpartir}
\usepackage{stmaryrd}
\usepackage{syntax}
\usepackage{tabto}
\usepackage{tikz}
\usepackage{url}
\usepackage{xspace}
\usepackage{comment}
\usepackage{framed}
\usepackage[clock]{ifsym}
\usepackage{breakurl}             
\input{scalalistings}

\newcommand{\ourtitle}{An Update on Deductive Synthesis and Repair in the Leon Tool}
\title{\ourtitle}
\author{Manos Koukoutos \qquad Etienne Kneuss  \qquad Viktor Kuncak
\institute{EPFL, Switzerland}
\email{firstname.lastname@epfl.ch}
}

\def\titlerunning{\ourtitle}
\def\authorrunning{Manos Koukoutos, Etienne Kneuss \& Viktor Kuncak}
\begin{document}
\maketitle

\begin{abstract}
  \input{abstract}
\end{abstract}

\input{introduction}

\input{reccalls}

\input{grammars}

\input{ste}

\input{table-eval}

\input{evaluation}

\input{related}

\bibliographystyle{eptcs}
\bibliography{managed,more}
\end{document}

%% file: scalalistings.tex
\lstdefinelanguage{ML}{
  alsoletter={*},
  morekeywords={datatype, of, if, *},
  sensitive=true,
  morecomment=[s]{/*}{*/},
  morestring=[b]"
}

\lstdefinelanguage{scala}{
  alsoletter={@,=,>},
  otherkeywords={passes},
  morekeywords={abstract, case, catch, choose, class, def, do, else, extends, false, final, finally, for, if, implicit, import, match, new, null, object, lazy, let,
override, package, private, protected, requires, return, sealed, super, this, throw, trait, try, true, type, val, var, while, with, yield, domain, 
postcondition, precondition, constraint, assert, forAll, _, return, @generator, ensure, require, ensuring, assuming, otherwise, asserting},
  sensitive=true,
  morecomment=[l]{//},
  morecomment=[s]{/*}{*/},
  morestring=[b]"
}

\newcommand{\codestyle}{\small\sffamily}

\newcommand{\SAND}{\mbox{\tt \&\&}\xspace}

\newcommand{\PP}{\mbox{\tt ++}\xspace}

\newcommand{\RA}{\Rightarrow}
\newcommand{\EQ}{\mbox{\tt ==}}
\newcommand{\NEQ}{\mbox{\tt !=}}

\newcommand{\SGT}{\mbox{\tt >}}

\makeatletter
\newcommand*\idstyle{%
        \expandafter\id@style\the\lst@token\relax
}
\def\id@style#1#2\relax{%
        \ifcat#1\relax\else
                \ifnum`#1=\uccode`#1
                \fi
        \fi
}
\makeatother

%% file: abstract.tex
We report our progress in scaling deductive synthesis and
repair of recursive functional Scala programs in the Leon tool.
We describe new techniques, including a more precise
mechanism for encoding the space of meaningful candidate
programs.  Our techniques increase the scope of 
synthesis by expanding the space of programs we
can synthesize and by reducing the synthesis time
in many cases.  As a new example, we present a run-length
encoding function for a list of values, which Leon can now
automatically synthesize from specification consisting of the decoding function
and the local minimality property of the encoded value.

%% file: introduction.tex
\section{Introduction}

This tool paper presents our recent improvements to deductive synthesis and repair
of the Leon tool
\cite{KneussETAL13SynthesisModuloRecursiveFunctions, 
KneussKoukoutosKuncak15DeductiveProgramRepair}.
The tool aims to synthesize (or repair) purely functional programs in a subset of Scala
containing mutually recursive functions. The generated code should provably satisfy a
specification, which is given by the programmer in the form of function pre- and postconditions
\cite{SuterKoeksalKuncak11SatisfiabilityModuloRecursivePrograms},
as well as (possibly symbolic) input-output examples 
\cite{KneussKoukoutosKuncak15DeductiveProgramRepair}.

Although our system does support interaction in synthesis \cite[Page 13]{Kneuss-Thesis}, the
evaluation in this paper focuses on fully automated synthesis, based
on searching a space of applicable rules. We employ a
set of deductive synthesis rules, that either decompose a synthesis
problem into simpler ones, or, if possible, solve it directly by
synthesizing a satisfying solution. The most notable closing rule is
Symbolic Term Exploration, which generates symbolic terms based on an
expression grammar.  The grammar is type-directed and
depends on the particular synthesis problem, e.g. it
produces constants and variables in scope, as well as calls
to available functions.

As demonstrated by our experimental evaluation,
our improvements allow the tool to synthesize
larger expressions, as well as to synthesize a wider variety of expressions.
This was made possible by refinements throughout the synthesis framework and its rules.
A notable novelty is a more general notion of program grammars whose non-terminals
are equipped with attributes.
These attributes enable us to produce certain types of
expressions in their normal form only and thus skip other
expressions that are syntactically different yet
semantically equivalent. We exploit for instance algebraic
laws for arithmetic operators. Such refined grammars may
thus prove useful for future versions of syntax-guided
synthesis format \cite{AlurETAL13SyntaxguidedSynthesis}.
By presenting a publicly available snapshot of our system and benchmarks
we hope to contribute to establishing a new
baseline for recursive program synthesis and repair.

The topic of deductive synthesis from specifications has
been explored actively over the past decades
\cite{DBLP:journals/toplas/MannaW80,
  Flener95LogicProgramSynthesisIncompleteInformation}.  A
key practical question that we aim to address is scalability
on program tasks containing recursive functions.  Most
existing systems require sketches, specifications of the
building block operators relevant for a given problem, or
definitions of domain-specific languages. Through such
additional specification users reduce the search space
compared to a more general case that our tool addresses.
The SyGuS synthesis competition
\cite{AlurETAL13SyntaxguidedSynthesis} helps objective
evaluation of synthesis tasks thanks to a grammar as an
explicit input. The benchmarks we discuss in this paper
require more expressive power than the current SyGuS
competition categories. We hope that in the future there
will be richer categories, and that we will also understand
good ways to leverage synthesis in restricted categories to
synthesize more complex programs. Here the situation is analogous
to verification systems that generate verification
conditions in SMT-LIB input format to prove correctness of
programs whose full semantics is beyond the abilities of SMT
solvers.

\subsection{Example}

\begin{figure}
\begin{minipage}[t]{0.5\linewidth}

\begin{lstlisting}
def decode[A](l: List[(BigInt, A)]): List[A] =
  l match {
    case Nil() $\RA$ Nil()
    case Cons((i, x), xs) $\RA$
      List.fill(i, x) $\PP$ decode(xs) }

def legal[A](l: List[(BigInt, A)]): Boolean =
  l match {
    case Nil() $\RA$ true
    case Cons((i, $\_$), Nil()) $\RA$ i $\SGT$ 0
    case Cons((i, x), tl@Cons(($\_$, y), $\_$)) $\RA$
      i $\SGT$ 0 && x $\NEQ$ y && legal(tl) }

def encode[A](l: List[A]): List[(BigInt, A)] =
  choose { res $\RA$
    legal(res) $\SAND$ decode(res) $\EQ$ l }
\end{lstlisting}
\end{minipage}
\begin{minipage}[t]{0.5\linewidth}
\begin{lstlisting}
def encode[A](l : List[A]): List[(BigInt, A)] = {
  l match {
    case Nil() $\RA$ Nil()
    case Cons(h0, t0) $\RA$
      val rec = encode(t0)
      rec match {
        case Nil() $\RA$ List((BigInt(1), h0))
        case Cons(h1 @ (h1$\_$1, h1$\_$2), t1) $\RA$
          if (h0 $\EQ$ h1$\_$2) {
            Cons((h1$\_$1 + BigInt(1), h1$\_$2), t1)
          } else {
            Cons((BigInt(1), h0), Cons(h1, t1)) 
          } } }
} ensuring { res $\RA$
  legal[A](res) $\SAND$ decode[A](res) $\EQ$ l
}
\end{lstlisting}
\end{minipage}
\caption{Run-length encoding problem (left) and solution (right) \label{example}}
\end{figure}

The left column of Figure \ref{example} specifies a run-length encoding algorithm.
Consider the \lstinline{encode} function specification.
It is expressed as a nondeterministic \lstinline{choose} construct,
which our tool will try to convert into a deterministic (executable) program
that satisfies the predicate within. This predicate expresses
\lstinline{encode} as the inverse of the \lstinline{decode} function that
generates \lstinline{legal} run-length encodings.
Leon is able to generate and formally verify the solution shown on the right
in about 20 seconds. It is one of the very few
tools that can solve similar problems with this level of automation.

The closest to the refinement of our 2013 approach that we here present 
is the recent \textsc{Synquid} system \cite{DBLP:conf/pldi/PolikarpovaKS16} that can 
synthesize both \lstinline{encode} and \lstinline{decode} functions
from a specification based on liquid types
in a very short
amount of time.
The version of the corresponding benchmark 
\footnote{\url{http://comcom.csail.mit.edu/demos/\#run-length}}
that was pointed to us, however, explicitly lists the zero
constant, successor, and predecessor function as the only
primitive building blocks for arithmetic expressions.  In
contrast, our system explores trees that, in addition to constants,
contain general binary arithmetic operations including addition and
substraction operators. As a result, 
our search space is notably larger.  In our attempts,
adding components corresponding to our search space made
the \textsc{Synquid} web example timeout after the 120 second
limit. Given that we are not experts in using
\textsc{Synquid}, a more systematic comparison remains to be
done in the future.

\subsection{Basic synthesis notation}

We will repeat a very brief overview of basic synthesis notation as given in
\cite{KneussKoukoutosKuncak15DeductiveProgramRepair}.
For more details, see \cite{KneussETAL13SynthesisModuloRecursiveFunctions}.

A synthesis problem is written as \br{\seqa}{\pcname}{\phi}{\seqx},
where \seqa are the input variables of the problem,
\pcname is the current path condition,
$\phi$ is the problem specification and
\seqx are the output variables.
\pcname is a function of \seqa,
whereas $\phi$ of both \seqa and \seqx.
A solution to a synthesis problem is a pair $\pg{\prename}{T}$,
where $T$ is the program term generated by synthesis,
and \prename is a precondition under which $T$ is a valid solution.
We illustrate the notation for decomposition rules
with a rule for splitting a problem containing a top-level
disjunction:
\begin{mathpar}
    \inferrule
  {
    \br{\seqa}{\pcname}{\phi_1}{\seqx} \vdash \pg{\prename_1}{T_1}
    \and
    \br{\seqa}{\pcname}{\phi_2}{\seqx} \vdash \pg{\prename_2}{T_2}
  }
  {
    \br{\seqa}{\pcname}{\phi_1 \lor \phi_2}{\seqx} \vdash \pg{\prename_1 \lor \prename_2}{\pgite{P_1}{T_1}{T_2}}
  }
\end{mathpar}

This rule should be interpreted as follows: from an input synthesis problem
$\br{\seqa}{\pcname}{\phi_1 \lor \phi_2}{\seqx}$, the rule decomposes
it in two subproblems: $\br{\seqa}{\pcname}{\phi_1}{\seqx}$ and
$\br{\seqa}{\pcname}{\phi_2}{\seqx}$. Given corresponding solutions
$\pg{\prename_1}{T_1}$ and $\pg{\prename_2}{T_2}$, the rule solves the initial
problem with $\pg{\prename_1 \lor \prename_2}{\pgite{P_1}{T_1}{T_2}}$.

%% file: reccalls.tex
\section{Recursive calls}

A key feature of Leon's synthesis is the ability to synthesize
programs with recursive function calls.

In \cite{KneussKoukoutosKuncak15DeductiveProgramRepair} we present a method
to introduce recursive calls that have good chance of not introducing non-termination.
Let us say we are trying to synthesize a function $foo$ with formal arguments $\overline{a}$.
Leon would track these arguments with a construct \terminates{foo(\overline{a})}.
The arguments $\overline{a}$ would then be transformed as needed by various decomposition rules.
When Symbolic Term Exploration is invoked, it will look for
the current \terminates{foo(\overline{a}')} construct,
and introduce recursive calls to $foo$ such that at least
one argument is smaller than initially, for a type-dependent definition of ``smaller''.
The rest of the arguments would be left free to be generated by symbolic term exploration.

Our new deployment of synthesis in Leon changes this approach, and instead uses a dedicated
deductive rule which introduces recursive calls to the synthesis context.
As before, it forces one argument of the function to be smaller,
but the rest of the arguments are fixed. 
This rule replaces all rules which introduced induction in any way.

Let $(\pcname \land a \leftarrow e)$ bind a fresh variable $a$ to the value $e$
in the path condition $\pcname$ of a problem.
Then the rule can be formally written as follows:

\[
    \inferrule[Introduce Rec. Calls]
  {
    \br{\seqa}{\pcname \land rec \leftarrow foo(a_1, \ldots, a'_i, \ldots, a_n)}{\phi_1}{\seqx} \vdash \pg{\prename_1}{T_1}
    \and
    a'_i \in argsSmaller(a_i, \pcname)
  }
  {
    \br{\seqa}{\terminates{foo(a_1, \ldots, a_i, \ldots, a_n)} \land \pcname }{\phi_1}{\seqx} 
       \vdash \pg{\prename_1}{T_1}
  }
\]

To define $argsSmaller$, let us consider an abstract class type $AC$
with a concrete descendant $CC$, and let $F$ be the fields of $CC$. Then
\[
\begin{array}{llll}
    &argsSmaller(i: Int,    & i > 0 \land \pcname) &= \{ i - 1 \}\\
    &argsSmaller(i: Int,    & i < 0 \land \pcname) &= \{ i + 1 \}\\
    &argsSmaller(i: BigInt, & i > 0 \land \pcname) &= \{ i - 1 \}\\
    &argsSmaller(i: BigInt, & i < 0 \land \pcname) &= \{ i + 1 \}\\
    &argsSmaller(c: AC,     & c: CC \land \pcname) &= \{ c.f\ \cup\ argsSmaller(c.f, \pcname)\ |\ f \in F \land f:AC \} \\
    &argsSmaller(v,         & \pcname            ) &= \emptyset \hskip 1cm  \text{otherwise}
\end{array}
\]

This approach cannot generate recursive calls where more than one argument changes;
for example, it cannot generate a recursive call to a function which updates an accumulator
while traversing a data structure.
However, it has the benefit that the variable $rec$,
which is bound to the result of the recursive call,
is now available to further decomposition rules.
This allows for new forms of programs to be synthesized.
For example, see the 6th line of the solution in Figure \ref{example}:
the introduced variable $rec$ is pattern-matched on by another rule,
which is necessary to solve the particular benchmark.

%% file: grammars.tex
\section{Term Grammars}
\label{sec:grammars}
\def\alt{~~|~~}

The main terminal rule of our framework, Symbolic Term Exploration,
generates symbolic terms with a context-free grammar.
The grammar takes into account the context of the current problem:
for example, it generates expressions containing variables in scope,
as well as calls to available functions.

Our previous implementation of expression grammars simply used
types as nonterminal symbols. For example, a grammar for integers
could be
$$
\gnt{Int} \gpo \gnt{Int} \gt{+} \gnt{Int} \alt
          \gnt{Int} \gt{-} \gnt{Int} \alt
          \gt{0} \alt
          \gt{a} \alt
          \gt{foo(}\gnt{Bool}\gt{)}
$$
where \lstinline{a: Int} is a parameter of the function under synthesis
and \lstinline{foo: Bool $\RA$ Int} is a function in scope.

However, such simple expression grammars have the disadvantage
of generating too many redundant terms.
One reason for that is that they are highly ambiguous.
For example, the above grammar would generate the term
\cl{a + a + foo(true)} in two different ways.
The other reason is that even syntactically distinct generated expressions
are very often semantically equivalent;
in our example, consider \hbox{\cl{a + foo(true)}} versus
\cl{foo(true) + a},
or \cl{a} versus \cl{a + 0} versus \cl{a + 0 + 0}.

Our current work addresses these issues by using a
richer representation of grammars. Nonterminal symbols are
enhanced with additional information beyond the type; we
refer to this contextual information as \emph{attributes}.
Attributes refine and filter production rules of an existing
grammar and enable us to fine-tune the shape of the
expression terms they represent.

In \cite{PerelmanETAL14TestdrivenSynthesis}, the authors apply a similar disambiguation
technique. In their case, the disambiguation happens after the terms
have been generated. In contrast, our attributes affect the
grammar itself, meaning that all terms produced are
automatically good candidates.
The bottom-up term generation technique used in \textsc{Transit} 
\cite{DBLP:conf/pldi/UdupaRDMMA13} 
merges even more equivalent 
expressions thanks to its evaluation-based under-approximation of
expression equivalence.

We now describe several of
the attributes defined in Leon and how they affect the
grammar productions. A generic production rule can be written as
\begin{equation}
    \label{gramprod}
    \gnt{T} \gpo \gt{f (}\gnt{T_1}, \gnt{T_2}, \ldots, \gnt{T_n}\gt{)}
\end{equation}
where $\textbf{f}$ is a function of the nonterminal symbols 
on the right-hand side of the rule.
The nonterminals $T$ and $T_i$ may
be plain types, or may already be annotated with attributes.
We represent attributes associated with a nonterminal in
braces.

\paragraph{Size and commutative operators.}
Iteratively generating bigger terms can be done by
gradually increasing the unfolding depth of the grammar.
This however causes the number of terms per depth to explode
double exponentially.
Instead, we use a
\emph{Sized} attribute that restricts the size of terms
produced\cite{Kneuss-Thesis}. For instance, $\gnt{Int}\gattr{|5|}$
produces only integer expressions of size 5, such as \cl{a +
b + c}.

Starting with a production rule of the form \ref{gramprod}
we can get the productions of $\gnt{T}\gattr{|s|}$ with
$$
\gnt{T}\gattr{|s|} \gpo \gt{f (}\gnt{T_1}\gattr{|s_1|}, \gnt{T_2}\gattr{|s_2|}, \ldots, \gnt{T_n}\gattr{|s_n|}\gt{)}
$$
for all combinations of $s_i > 0$ such that $\sum{s_i} = s - size(f)$,
where $size(f)$ a cost we associate with $f$.
If $n = 0$, the production is kept only if $size(f) = s$.
Additionally, if $f$ is a commutative operator,
we require that $\forall i < j.~ s_i \ge s_j$.  As a result,
only left-heavy terms are produced by the grammar (i.e. \cl{(a
* b) + c} and not the equivalent \cl{c + (a * b)}).

\paragraph{Associative operators.}

\def\notf{\neg f}

To remove redundancy caused by operator associativity,
we require that all associative operators associate to the left.
Let $\textbf{f}$ be an associative operator and $\notf$ an attribute
that disallows production rules with operator $\textbf{f}$.
Then $\gnt{T} \gpo \gt{f (}\gnt{T_1}, \gnt{T_2}\gt{)}$
becomes $\gnt{T} \gpo \gt{f (}\gnt{T_1}, \gnt{T_2}\gattr{\notf}\gt{)}$,
and rules of the form \hbox{$\gnt{T}\gattr{\notf} \gpo \gt{f
(}\gnt{T_1}, \gnt{T_2} \gt{)}$}
are removed.

\paragraph{Ground Terms.}
\label{ground}
\def\gr{G}
\def\notgr{\neg G}
\newcommand\gri[1]{G_{#1}}

We want to avoid that our grammars generate ground terms for two reasons:
firstly, because different combinations of ground terms may end up
simplifying to equivalent programs (consider \cl{1 + 3} and \cl{2 + 2});
secondly, because our system includes another dedicated rule
which is much more efficient than Symbolic Term Exploration
in discovering ground terms.

Let the attribute $\gr$ denote that a ground term is expected, whereas $\notgr$ that a ground term
is disallowed. Then
\[
\begin{array}{rl}
    \gnt{T}\gattr{\gr}    \gpo& \gt{f (}\gnt{T_1}\gattr{\gr}, \gnt{T_2}\gattr{\gr}, \ldots, \gnt{T_n}\gattr{\gr}\gt{)}\\
    \gnt{T}\gattr{\notgr} \gpo& \gt{f (}\gnt{T_1}\gattr{\gri{1}}, \gnt{T_2}\gattr{\gri{2}}, \ldots, \gnt{T_n}\gattr{\gri{n}}\gt{)}
\end{array}
\]
for all combinations of $G_i \in \{ \gr, \notgr \}$ such that at least one $\gri{i}$ is $\notgr$.
If $\gt{c}$ is a constant and $\gt{v}$ is a variable,
rules of the forms \hbox{$\gnt{T}\gattr{\notgr} \gpo \gt{c}$} and $\gnt{T}\gattr{\gr} \gpo \gt{v}$
are removed.

The $\notgr$ attribute is attached to the top-level symbol of the grammar.

\paragraph{Neutral elements.}
\def\nz{\neg 0}

Several arithmetic operators have so-called neutral or
absorbing operands that are sources of redundancies.
For example, terms such as \cl{e + 0}, \cl{e / 1}, or
\cl{e * 0} are all equivalent to a shorter form.
We eliminate these from the grammar by using an attribute
that excludes neutral elements: $\gnt{Int} \gpo \gnt{Int} \gt{+} \gnt{Int}$ becomes
$\gnt{Int} \gpo \gnt{Int}\gattr{\nz} \gt{+} \gnt{Int}\gattr{\nz}$. Naturally,
the production $\gnt{Int} \gpo \gt{0}$ is excluded from
$\gnt{Int}\gattr{\nz}$.

%% file: ste.tex
\section{Symbolic Term Exploration}

\newcommand{\fcall}[1]{\textsc{#1}}

\input{algorithm}

The main closing rule that our tool employs is Symbolic Term Exploration.
Although the algorithm has not changed much conceptually since previously presented
in \cite{KneussETAL13SynthesisModuloRecursiveFunctions},
our implementation has matured as we gained experience using it.
In this section we document our current design choices.
Additionally, we provide detailed pseudocode for our approach,
hoping it will serve as a starting point for similar implementations.

Symbolic Term Exploration (STE) unfolds a grammar as described in the Section \ref{sec:grammars}
to create a set of candidate programs, which are represented all together
with a symbolic program tree \cite{KneussETAL13SynthesisModuloRecursiveFunctions}.
These programs are first filtered with concrete execution based on a set of tests.
The ones that survive concrete testing have to be handled symbolically
with the aid of a Leon solver \cite{SuterKoeksalKuncak11SatisfiabilityModuloRecursivePrograms, suter-thesis},
which reduces verification of a subset of Scala to a stream of queries 
for SMT solvers (currently Z3 \cite{MouraBjoerner08Z3EfficientSmtSolver}
and CVC4 \cite{DetersETAL14TourCvc4HowItWorksHow}).

Our synthesis tool uses two approaches to determine if the representation of a set of
programs contains a valid program:
(1) if the number of remaining programs is relatively small, we try to prove
or disprove each one separately with the Leon solver. If a program is
proven correct, we have a satisfactory solution; otherwise, the solver
generates a fresh counterexample which we add to our test base,
as it may help exclude further programs.
(2) if the number of remaining programs is large, we query the solver for a program
that satisfies the specification for at least one input. If no such program
exists, then no program in our candidate set satisfies the specification,
and STE fails; otherwise, we hope that this program has good chances to
be a satisfactory solution, and we try to prove it valid as in (1).

Our experience using STE since the previous iterations of the tool has shown that,
even for thousands of programs and tens of tests,
concrete execution is usually faster than a single SMT query.
An explanation is that many of our tests are generated by automatic
data generators, and they tend to be quite small (small numeric values or
data structures of few nodes). For that reason, we have adapted our implementation
to rely as much as possible on concrete execution with the following adjustments:
\begin{itemize}
    \item We concretely test candidate programs against every counterexample
        as soon as it is discovered by the Leon solver.
    \item We make sure to utilize parts of the solution that have already been
        discovered by other deductive rules \cite{Kneuss-Thesis}. From these parts we construct
        a partial solution, with a placeholder in place of the current problem.
        For example, consider we are trying to solve the example of Figure \ref{example}.
        After case-splitting on \lstinline{l} and solving the \lstinline{Nil} case,
        the partial solution would be
\begin{lstlisting}
l match {
  case Nil() $\RA$ Nil()
  case Cons(h0, t0) $\RA$ ???
}
\end{lstlisting}

        This expression will temporarily replace the original implementation
        of \lstinline{encode} (the \lstinline{choose}) in the program.
        During concrete execution and validation, the placeholder \lstinline{???}
        is set each time to the program we are currently testing.
        Similarly, during the discovery of a tentative program,
        the placeholder is set to the tree representation of the set
        of available programs.

    \item Another simple heuristic we introduced that yields good results in practice
        is to sort available tests according to the number of programs that
        failed on them. This way, tests that have been more successful in the past
        in excluding programs get executed first.
\end{itemize}

Algorithm \ref{ste} shows an overview of our current STE implementation. 
The main function, \fcall{Ste},
takes as arguments a synthesis problem, an expression grammar G
and a desired maximum size of generated programs. It uses auxiliary
functions \fcall{Validate} and \fcall{ConcreteTest} shown
below, as well as two global variables $I$ and $P$.

%% file: algorithm.tex
\begin{algorithm}[t]
\caption{Symbolic Term Exploration}
\label{ste}
\begin{algorithmic}
\State \textbf{var} $I=$ Initial list of examples
\State \textbf{var} $P = \emptyset$                         \Comment Set of examples represented as a tree
~\\
\Function{Ste}{\br{\seqa}{\pcname}{\phi}{\seqx}, G, $maxSize$}
    \For{$n \gets 1$ \textbf{to} $maxSize$}                 \Comment{STE loop over program sizes}
        \State $P=$ \fcall{Unfold}(G, $n$)                  \Comment{Generate programs of size $n$ from grammar G}
        \State \fcall{ConcreteTest}(I, $\phi$)              \Comment Exclude programs by concrete execution
        \While{$P \ne\emptyset$}                            \Comment Main STE Loop
            \If{$\vert P\vert$ sufficiently reduced}
                \ForAll{$p \in P$}                          \Comment{Try to validate individually}
                    \If {\fcall{Validate}(\br{\seqa}{\pcname}{\phi}{\seqx}, p)}
                        \Return p                           \Comment{Found valid solution!}
                    \EndIf
                \EndFor
            \EndIf
            \State \textbf{let} $f = (\pcname \land p \in P \land \phi[\seqx / p(\seq{a}')])$, $\seq{a}'$ fresh
            \If{$\neg \fcall{LeonSolverSat}(f)$}                   \Comment No program of this size works
                \State Break to the next value of $n$
            \Else
                \State \textbf{let} $p_0 = \fcall{LeonSolverModel}(f)$
                \If {\fcall{Validate}(\br{\seqa}{\pcname}{\phi}{\seqx}, $p_0$)}
                    \State \Return $p_0$                    \Comment Found valid solution!
                \EndIf
            \EndIf
        \EndWhile
    \EndFor
    \Return \textsc{Fail}                                   \Comment No program found for any program size
\EndFunction
\end{algorithmic}
~\\
\begin{minipage}[t]{.5\textwidth}
\begin{algorithmic}
\Function{Validate}{\br{\seqa}{\pcname}{\phi}{\seqx}, p}
    \State \textbf{let} $f = (\pcname \land \neg \phi[\seqx / p(\seqa)])$
    \If{$\neg \fcall{LeonSolverSat}(f)$}
        \Return true
    \Else
        \State $P= P \setminus \{p\}$
        \State \textbf{let} $a_0 =$ \fcall{LeonSolverModel}($f$)
        \State $\fcall{ConcreteTest}(\{a_0\}, \phi)$
        \State $I= I \cup \{a_0\}$
        \State \Return false
    \EndIf
\EndFunction
\end{algorithmic}
\end{minipage}
\begin{minipage}[t]{.5\textwidth}
\begin{algorithmic}
\Procedure{ConcreteTest}{J, $\phi$}
    \ForAll{$p \in P$}
        \ForAll{$\seq{a_0} \in J$}
            \If{$\neg \fcall{Execute}(\phi[\seqx / p(\seq{a_0})])$}
                \State \fcall{IncreasePriority}(J, $\seq{a_0}$)
                \State $P= P \setminus \{p\}$
            \EndIf
        \EndFor
    \EndFor
\EndProcedure
\end{algorithmic}
\end{minipage}

\end{algorithm}

%% file: table-eval.tex
\newcommand{\leg}[2]{{#1: #2\newline }}
\newcommand{\mc}[2]{\multicolumn{2}{#1}{#2}}
\newcommand{\tme}{\showclock{0}{15}}
\renewcommand{\bname}[1]{{\footnotesize \texttt{#1}}}
\newcolumntype{R}{>{\hspace{-7pt}}r}
\newcolumntype{C}{>{\hspace{-3pt}}c}

\begin{table}[t]
    \begin{center}
    \begin{tabular}{l|rr|CR|CR|CR|CR|CR}
        Operation                        & \mc{c|}{Sizes} & \mc{c|}{Prev}    & \mc{c|}{Size = 7}    & \mc{c|}{STE}          & \mc{c|}{Rec}      & \mc{c}{TG}         \\
                                         &  Pr  &  Sol   & $\vdash$  & \tme  & $\vdash$  & \tme     & $\vdash$   & \tme     & $\vdash$  & \tme  & $\vdash$   & \tme  \\
        \hline                                                                                                                                                       
        \bname{BatchedQueue.enqueue}     &   92 &   26   &       X   &       & ($\surd$) &   22.8   &  ($\surd$) &  18.7    & ($\surd$) & 18.8  &  ($\surd$) &  15.4 \\
        \bname{List.split     }          &   84 &   33   &       X   &       &       X   &          &  X         &          & $\surd$   & 2.4   &  $\surd$   &  2.5  \\
        \bname{AddressBook.make}         &   43 &   36   &       X   &       &       X   &          &  X         &          & $\surd$   & 4.2   &  $\surd$   &  4.0  \\
        \bname{RunLength.encode}         &  118 &   39   &       X   &       &       X   &          &  X         &          & $\surd$   & 18.7  &  $\surd$   &  20.3 \\
        \bname{Diffs.diffs}              &   63 &   24   &       X   &       &       X   &          &  X         &          & $\surd$   & 24.5  &  $\surd$   &  11.8 \\
        \hline
        \bname{List.insert    }          &   61 &    3   & $\surd$   & 0.9   & $\surd$   & 0.6      &  $\surd$   &   0.8    & $\surd$   & 0.8   &  $\surd$   &  0.7  \\
        \bname{List.delete   }           &   63 &   19   & $\surd$   & 4.2   & $\surd$   & 39.7     &  $\surd$   &   12.6   & $\surd$   & 12.1  &  $\surd$   &  8.6  \\
        \bname{List.union      }         &   77 &   12   & $\surd$   & 7.9   & $\surd$   & 13.7     &  $\surd$   &   3.7    & $\surd$   & 3.6   &  $\surd$   &  2.6  \\
        \bname{List.diff    }            &  109 &   12   & $\surd$   & 6.4   & $\surd$   & 110.9    &  $\surd$   &   23.2   & $\surd$   & 24.7  &  $\surd$   &  12.4 \\
        \bname{List.listOfSize }         &   38 &   11   & $\surd$   & 1.4   & $\surd$   & 1.7      &  $\surd$   &   1.1    & $\surd$   & 1.6   &  $\surd$   &  1.4  \\
        \bname{SortedList.insert}        &   94 &   30   & ($\surd$) & 18.0  & ($\surd$) & 125.0    &  ($\surd$) &   17.5   & ($\surd$) & 24.3  &  ($\surd$) &  16.0 \\
        \bname{SortedList.insertAlways}  &  108 &   32   & $\surd$   & 22.8  & $\surd$   & 139.3    &  $\surd$   &   32.6   & $\surd$   & 35.2  &  $\surd$   &  21.0 \\
        \bname{SortedList.delete   }     &   94 &   19   & ($\surd$) & 7.6   & ($\surd$) & 57.6     &  ($\surd$) &   19.8   & ($\surd$) & 16.8  &  ($\surd$) &  15.8 \\
        \bname{SortedList.union  }       &  142 &   12   & $\surd$   & 7.5   & $\surd$   & 12.7     &  $\surd$   &   4.1    & $\surd$   & 4.5   &  $\surd$   &  3.3  \\
        \bname{SortedList.diff }         &  140 &   12   & $\surd$   & 5.8   & $\surd$   & 104.0    &  $\surd$   &   12.4   & $\surd$   & 13.6  &  $\surd$   &  6.7  \\
        \bname{SortedList.insertionSort} &  129 &   11   & $\surd$   & 1.4   & $\surd$   & 2.7      &  $\surd$   &   2.5    & $\surd$   & 2.4   &  $\surd$   &  1.7  \\
        \bname{StrictSortedList.insert } &   94 &   30   & $\surd$   & 13.4  & $\surd$   & 111.5    &  $\surd$   &   16.7   & $\surd$   & 24.9  &  $\surd$   &  15.8 \\
        \bname{StrictSortedList.delete}  &   94 &   19   & ($\surd$) & 9.1   & ($\surd$) & 64.9     &  ($\surd$) &   22.3   & $\surd$   & 19.9  &  $\surd$   &  15.5 \\
        \bname{StrictSortedList.union}   &  142 &   12   & $\surd$   & 7.7   & $\surd$   & 12.8     &  $\surd$   &   4.3    & $\surd$   & 4.6   &  $\surd$   &  3.1  \\
        \bname{UnaryNumerals.add }       &   46 &   10   & $\surd$   & 4.6   & $\surd$   & 4.3      &  $\surd$   &   3.1    & $\surd$   & 3.8   &  $\surd$   &  3.1  \\
        \bname{UnaryNumerals.distinct}   &   71 &    4   & $\surd$   & 2.2   & $\surd$   & 2.1      &  $\surd$   &   2.0    & $\surd$   & 2.2   &  $\surd$   &  2.0  \\
        \bname{UnaryNumerals.mult}       &   46 &   11   & $\surd$   & 4.6   & $\surd$   & 10.7     &  $\surd$   &   6.9    & $\surd$   & 5.6   &  $\surd$   &  5.7  \\
        \bname{BatchedQueue.dequeue}     &   68 &   12   & ($\surd$) & 13.0  & ($\surd$) & 10.2     &  ($\surd$) &   21.5   & ($\surd$) & 21.3  &  ($\surd$) &  17.8 \\
        \bname{AddressBook.merge}        &  104 &   17   & ($\surd$) & 6.0   & ($\surd$) & 7.4      &  ($\surd$) &   18.7   & ($\surd$) & 19.0  &  ($\surd$) &  17.6 \\
    \end{tabular}
    \end{center}
    \caption{Benchmarks for synthesis \label{eval-synth}}
\end{table}

\begin{table}
\begin{center}
\begin{tabular}{l|rr|rr|rr}
Operation                   & \mc{c|}{Sizes}  & \mc{c|}{Previous}  & \mc{c}{Current} \\
                            & Prog &  Sol     & Test & Repair      & Test & Repair   \\
\hline
\bname{ Compiler.desugar  } & 670  &    3     &  1.0 &   1.9       & 0.8 &  3.1 \\
\bname{ Compiler.desugar  } & 668  &    2     &  0.9 &  12.3       & 0.8 &  4.5 \\
\bname{ Compiler.desugar  } & 672  &    7     &  0.6 &   1.4       & 5.3 &  1.5 \\
\bname{ Compiler.desugar  } & 672  &    7     &  1.1 &   1.5       & 0.7 &  2.7 \\
\bname{ Compiler.desugar  } & 672  &   14     &  1.0 &  12.8       & 0.8 &  2.7 \\
\bname{ Compiler.simplify } & 718  &    4     &  0.6 &   1.4       & 0.4 &  2.5 \\
\bname{ Compiler.simplify } & 718  &    2     &  0.6 &   1.4       & 0.4 &  1.1 \\
\bname{ Heap.merge        } & 341  &    3     &  1.4 &   2.7       & 2.5 & 14.2 \\
\bname{ Heap.merge        } & 341  &    1     &  0.7 &   1.3       & 2.1 &  2.2 \\
\bname{ Heap.merge        } & 341  &    3     &  1.4 &   2.6       & 2.6 & 12.6 \\
\bname{ Heap.merge        } & 341  &    9     &  1.0 &   2.2       & 2.4 & 10.0 \\
\bname{ Heap.merge        } & 343  &    5     &  0.9 &   2.6       & 2.3 & 12.0 \\
\bname{ Heap.merge        } & 341  &    2     &  1.1 &  13.6       & 2.1 &  3.3 \\
\bname{ Heap.insert       } & 304  &    8     &  4.3 &   1.0       & 2.9 &  6.0 \\
\bname{ Heap.makeN        } & 343  &    7     &  2.0 &   1.3       & 1.1 &  8.5 \\
\bname{ List.pad          } & 802  &    8     &  0.7 &   1.2       & 1.0 &  2.7 \\
\bname{ List.++           } & 712  &    3     &  1.9 &   1.0       & 1.2 &  4.8 \\
\bname{ List.:+           } & 744  &    1     &  1.5 &   1.0       & 0.7 &  1.5 \\
\bname{ List.replace      } & 746  &    6     &  1.2 &  10.4       & 1.0 &  5.9 \\
\bname{ List.count        } & 799  &    3     &  0.7 &   1.3       & 1.6 &  9.7 \\
\bname{ List.find         } & 799  &    2     &  2.9 &   3.5       & 1.5 &  9.5 \\
\bname{ List.find         } & 801  &    4     &  2.6 &   3.6       & 1.6 &  9.7 \\
\bname{ List.find         } & 802  &    4     &  4.4 &   5.2       & 0.9 & 25.5 \\
\bname{ List.size         } & 748  &    4     &  1.5 &   1.0       & 0.7 &  1.5 \\
\bname{ List.sum          } & 746  &    4     &  1.1 &   1.3       & 0.5 &  1.6 \\
\bname{ List.-            } & 746  &    1     &  1.1 &   1.0       & 2.4 &  9.7 \\
\bname{ List.drop         } & 787  &    4     &  1.3 &  16.6       &   X &   X  \\
\bname{ Numerical.power   } & 172  &    5     &  0.2 &   1.0       & 0.2 &  3.8 \\
\bname{ Numerical.moddiv  } & 121  &    3     &  0.2 &   0.8       & 0.1 &  1.3 \\
\bname{ MergeSort.split   } & 228  &    5     &  1.8 &   2.8       & 1.6 &  6.6 \\
\bname{ MergeSort.merge   } & 230  &    7     &  1.3 &   1.2       & 1.8 &  2.9 \\
\bname{ MergeSort.merge   } & 230  &    3     &  1.2 &   1.7       & 1.6 &  5.1 \\
\bname{ MergeSort.merge   } & 228  &    5     &  1.1 &   1.2       & 1.8 &  3.0 \\
\bname{ MergeSort.merge   } & 230  &    1     &  1.4 &  20.9       & 1.5 &  1.5 \\
\end{tabular}
\caption{Benchmarks for repair \label{eval-repair}}
\end{center}
\end{table}

%% file: evaluation.tex
\section{Evaluation}

We evaluated the improvements presented in the previous sections against
the previous version of Leon and we present the results in Table \ref{eval-synth}.

The first column of Table \ref{eval-synth} gives an indication of the difficulty of each benchmark:
\textit{Prog} indicates the total size of the program in AST nodes, and
\textit{Sol} indicates the size of the solution generated by the latest version of Leon.
For each version of Leon tested and each benchmark, we list the running time as well as whether the tool
produced and verified a solution ($\surd$), failed altogether (X),
or produced a solution but could not verify it ($\surd$ in parentheses).
An X means synthesis failed for this benchmark; either the benchmark timed out after 200 seconds,
or the synthesizer exhausted its search space without coming up with a solution.

To make the effect of each individual improvement clearer,
the following columns of the table showcase the performance of
the tool as various features are added.
The first column presents the original version of Leon.
In that version, we constrained Symbolic Term Exploration
to expressions of size up to 5.
However, after the latest optimizations we found it is viable
to increase the bound to 7. This immediately solves one additional benchmark,
but is not viable by itself due to the large slowdowns it introduces.
The results for this configuration are presented for completeness 
under the column \textit{Size = 7}.
The next columns introduce respectively
the new improved Symbolic Term Exploration (\textit{STE}),
the new rule for recursive functions (\textit{Rec})
and finally the optimized term grammars (\textit{TG}).

\paragraph{Observations:}
The new version of the tool was able to solve
five new hard benchmarks which were out of scope for the previous versions.
It also produces a much more concise, and thus verifiable,
solution for an additional benchmark (\bname{StrictSortedList.delete}).
Some of the easier benchmarks do present some slowdown.
This is mostly caused by failing STE instances that need to
exhaust a larger search space. We believe this is a small
price to pay: Our focus is to push the limits of what can be
synthesized in a reasonable amount of time, rather than to
optimize for simpler benchmarks.

Concerning individual improvements, we can see that the STE improvements
greatly improve the performance of the tool (compared to the version with the same STE size,
of course), without solving additional benchmarks.
The improved term grammars also have a significant, if smaller,
effect on running times; however, we do expect the improvement to
become more significant as our tool scales to larger expression sizes
due to the exponential nature of the problem.
Finally, the new rule for recursive calls does not improve running times,
but extends the search space of the tool and thus solves an additional
four benchmarks.

Table \ref{eval-repair} displays results for a set of benchmarks for program repair
identical to the one presented in \cite{KneussKoukoutosKuncak15DeductiveProgramRepair}.
The results here are not so interesting, so we just include the times for
the initial and final versions of Leon with all the optimizations.
We can see that the benchmarks generally show some delay relatively to the previous version.
This is mostly due to changes that increase the system's reliability,
with some cost in performance (more robust nondeterministic evaluator,
support for multiple synthesis solutions etc.)
Additionally, one benchmark is not solvable any more due to the change in
handling of recursive functions. However, we could arguably revert to grammar-generated
recursive calls for repair benchmarks, as required recursive calls are likely to be
present in the program already (of course, they would be subject to repair).

The version of Leon used for evaluation can be found at
\url{https://github.com/epfl-lara/leon/tree/synt2016},
and the benchmarks at
\url{https://github.com/epfl-lara/leon/tree/synt2016/testcases/synt2016}.

%% file: related.tex
\section{Related work}

Other recent tools that focus on deductive synthesis of recursive programs
from formal specification include \textsc{SyntRec} \cite{DBLP:journals/corr/InalaQLS15},
\textsc{Synapse} \cite{BornholtETAL16OptimizingSynthesisMetasketches}
and \textsc{Synquid} \cite{DBLP:conf/pldi/PolikarpovaKS16}.
\textsc{SyntRec} and \textsc{Synapse} use a similar approach based on user-defined 
\textit{generators} (or metasketches) that describe high-level, reusable patterns of computation,
in the spirit of \textsc{Sketch} \cite{sketch}.
The programmer interacts with the system by providing an appropriate generator
for the task at hand, which is then used by the system to synthesize a complete program.
\textsc{SyntRec} validates candidate program with bounded checking,
whereas \textsc{Synapse} uses SMT.
These approaches scale better for some benchmarks, but require the programmer
to have significant insight into the form of the resulting program.
In \textsc{Synquid}, the target specification is given in the form of a liquid type \cite{liquid}.
Additionally, the user provides the set of usable program components.
The authors modify the liquid type inference algorithm to enable top-down breakdown of
a liquid type, and use the inference rules as deductive synthesis rules.
Conditionals are generated with a form of condition abduction.
Compared to Leon, \textsc{Synquid} specifications tend to be much longer and require more insight,
as the programmer needs to provide the liquid type signatures of all intermediate components
used by the synthesizer.

A large body of research in the area has focused on inductive synthesis,
or synthesis from input/output examples.
Examples are a more intuitive form of specification,
especially for non-expert users, and can be reasoned about
with concrete execution rather than formal proofs.
However, examples can never fully specify the intention of the programmer
for an infinite domain, leading to ambiguities in the resulting synthesized programs.
\textsc{Esher} \cite{AlbarghouthiGulwaniKincaid13RecursiveProgramSynthesis} and
LaSy \cite{PerelmanETAL14TestdrivenSynthesis}
use a set of input/output examples and a set of program components
to automatically synthesize progressively more complicated code snippets,
until one is discovered which satisfies all input-output pairs.
In \cite{FrankleETAL16ExampledirectedSynthesisTypetheoreticInterpretation},
the authors use a type-based approach,
where an input/output example is viewed as a singleton refinement type. 
A solution is satisfactory if its type is a supertype of all provided examples.
AutoFix \cite{PeiETAL15AutomatedProgramRepairIntegratedDevelopmentEnvironment,
wei2010automated} locates and fixes bugs in imperative Eiffel code
decorated with formal contracts.
Suspicious statements are located based on their presence in passing and failing
example traces.
In \cite{PolozovGulwani15FlashmetaFrameworkInductiveProgramSynthesis} the authors define
a generic framework for synthesis-by-example. The framework provides a fixed synthesis algorithm
and can be instantiated with a specific DSL, along with weights for its expressions
and other domain-specific knowledge.
The synthesizer is in dialogue with the programmer to eliminate ambiguities
in the generated programs.

Finally, a direction of work has been synthesizing snippets that interact with APIs.
Since large APIs are an integral part of programming,
the focus of this work is shifted to higher-level code that is mostly restricted to a series
of API calls as opposed to application of primitive operations.
These tools usually require a corpus of code in the target language
to construct a language model offline,
from which they extract weights which guide the synthesis algorithm.
Reinking and Piskac \cite{ReinkingPiskac15TypedirectedApproachToProgramRepair}
focus on repair of type-incorrect API invocations.
The line of work of Gvero et al. \cite{
GveroKuncak15SynthesizingJavaExpressionsFromFreeformQueries,
GveroETAL13CompleteCompletionUsingTypesWeights,
DBLP:conf/cav/GveroKP11}
aims to synthesize queries to APIs in Scala/Java within an IDE environment,
using the local environment at the point of invocation of the tool,
(including local variables and API functions),
or, more recently, taking a free form query as input.
In \cite{DBLP:conf/pldi/PerelmanGBG12}, the input to the synthesizer is a partial expression,
which can encode calls to an unknown function on known arguments
or, given an object, an invocation of an unknown method or lookup of an unknown field of that object.
A synthesis algorithm completes those partial expressions to obtain a complete program.

%% file: main.bbl
\begin{thebibliography}{10}
\providecommand{\bibitemdeclare}[2]{}
\providecommand{\surnamestart}{}
\providecommand{\surnameend}{}
\providecommand{\urlprefix}{Available at }
\providecommand{\url}[1]{\texttt{#1}}
\providecommand{\href}[2]{\texttt{#2}}
\providecommand{\urlalt}[2]{\href{#1}{#2}}
\providecommand{\doi}[1]{doi:\urlalt{http://dx.doi.org/#1}{#1}}
\providecommand{\bibinfo}[2]{#2}

\bibitemdeclare{inproceedings}{AlbarghouthiGulwaniKincaid13RecursiveProgramSynthesis}
\bibitem{AlbarghouthiGulwaniKincaid13RecursiveProgramSynthesis}
\bibinfo{author}{Aws \surnamestart Albarghouthi\surnameend},
  \bibinfo{author}{Sumit \surnamestart Gulwani\surnameend} \&
  \bibinfo{author}{Zachary \surnamestart Kincaid\surnameend}
  (\bibinfo{year}{2013}): \emph{\bibinfo{title}{Recursive Program Synthesis}}.
\newblock In \bibinfo{editor}{Natasha \surnamestart Sharygina\surnameend} \&
  \bibinfo{editor}{Helmut \surnamestart Veith\surnameend}, editors: {\sl
  \bibinfo{booktitle}{CAV}}, {\sl \bibinfo{series}{LNCS}}
  \bibinfo{volume}{8044}, \bibinfo{publisher}{Springer}, pp.
  \bibinfo{pages}{934--950}, \doi{10.1007/978-3-642-39799-8_67}.

\bibitemdeclare{inproceedings}{AlurETAL13SyntaxguidedSynthesis}
\bibitem{AlurETAL13SyntaxguidedSynthesis}
\bibinfo{author}{Rajeev \surnamestart Alur\surnameend},
  \bibinfo{author}{Rastislav \surnamestart Bod{\'{\i}}k\surnameend},
  \bibinfo{author}{Garvit \surnamestart Juniwal\surnameend},
  \bibinfo{author}{Milo M.~K. \surnamestart Martin\surnameend},
  \bibinfo{author}{Mukund \surnamestart Raghothaman\surnameend},
  \bibinfo{author}{Sanjit~A. \surnamestart Seshia\surnameend},
  \bibinfo{author}{Rishabh \surnamestart Singh\surnameend},
  \bibinfo{author}{Armando \surnamestart Solar{-}Lezama\surnameend},
  \bibinfo{author}{Emina \surnamestart Torlak\surnameend} \&
  \bibinfo{author}{Abhishek \surnamestart Udupa\surnameend}
  (\bibinfo{year}{2013}): \emph{\bibinfo{title}{Syntax-guided synthesis}}.
\newblock In: {\sl \bibinfo{booktitle}{FMCAD}}, \bibinfo{publisher}{{IEEE}},
  pp. \bibinfo{pages}{1--8}, \doi{10.3233/978-1-61499-495-4-1}.

\bibitemdeclare{inproceedings}{BornholtETAL16OptimizingSynthesisMetasketches}
\bibitem{BornholtETAL16OptimizingSynthesisMetasketches}
\bibinfo{author}{James \surnamestart Bornholt\surnameend},
  \bibinfo{author}{Emina \surnamestart Torlak\surnameend}, \bibinfo{author}{Dan
  \surnamestart Grossman\surnameend} \& \bibinfo{author}{Luis \surnamestart
  Ceze\surnameend} (\bibinfo{year}{2016}): \emph{\bibinfo{title}{Optimizing
  synthesis with metasketches}}.
\newblock In \bibinfo{editor}{Rastislav \surnamestart Bod{\'{\i}}k\surnameend}
  \& \bibinfo{editor}{Rupak \surnamestart Majumdar\surnameend}, editors: {\sl
  \bibinfo{booktitle}{POPL}}, \bibinfo{publisher}{{ACM}}, pp.
  \bibinfo{pages}{775--788}, \doi{10.1145/2837614.2837666}.

\bibitemdeclare{inproceedings}{DetersETAL14TourCvc4HowItWorksHow}
\bibitem{DetersETAL14TourCvc4HowItWorksHow}
\bibinfo{author}{Morgan \surnamestart Deters\surnameend},
  \bibinfo{author}{Andrew \surnamestart Reynolds\surnameend},
  \bibinfo{author}{Tim \surnamestart King\surnameend},
  \bibinfo{author}{Clark~W. \surnamestart Barrett\surnameend} \&
  \bibinfo{author}{Cesare \surnamestart Tinelli\surnameend}
  (\bibinfo{year}{2014}): \emph{\bibinfo{title}{A tour of {CVC4:} How it works,
  and how to use it}}.
\newblock In: {\sl \bibinfo{booktitle}{FMCAD}}, \bibinfo{publisher}{{IEEE}},
  p.~\bibinfo{pages}{7}, \doi{10.1109/FMCAD.2014.6987586}.

\bibitemdeclare{book}{Flener95LogicProgramSynthesisIncompleteInformation}
\bibitem{Flener95LogicProgramSynthesisIncompleteInformation}
\bibinfo{author}{Pierre \surnamestart Flener\surnameend}
  (\bibinfo{year}{1995}): \emph{\bibinfo{title}{Logic Program Synthesis from
  Incomplete Information}}.
\newblock \bibinfo{publisher}{Springer}, \doi{10.1007/978-1-4615-2205-8}.

\bibitemdeclare{inproceedings}{FrankleETAL16ExampledirectedSynthesisTypetheoreticInterpretation}
\bibitem{FrankleETAL16ExampledirectedSynthesisTypetheoreticInterpretation}
\bibinfo{author}{Jonathan \surnamestart Frankle\surnameend},
  \bibinfo{author}{Peter{-}Michael \surnamestart Osera\surnameend},
  \bibinfo{author}{David \surnamestart Walker\surnameend} \&
  \bibinfo{author}{Steve \surnamestart Zdancewic\surnameend}
  (\bibinfo{year}{2016}): \emph{\bibinfo{title}{Example-directed synthesis: a
  type-theoretic interpretation}}.
\newblock In \bibinfo{editor}{Rastislav \surnamestart Bod{\'{\i}}k\surnameend}
  \& \bibinfo{editor}{Rupak \surnamestart Majumdar\surnameend}, editors: {\sl
  \bibinfo{booktitle}{POPL}}, \bibinfo{publisher}{{ACM}}, pp.
  \bibinfo{pages}{802--815}, \doi{10.1145/2837614.2837629}.

\bibitemdeclare{inproceedings}{GveroKuncak15SynthesizingJavaExpressionsFromFreeformQueries}
\bibitem{GveroKuncak15SynthesizingJavaExpressionsFromFreeformQueries}
\bibinfo{author}{Tihomir \surnamestart Gvero\surnameend} \&
  \bibinfo{author}{Viktor \surnamestart Kuncak\surnameend}
  (\bibinfo{year}{2015}): \emph{\bibinfo{title}{Synthesizing Java expressions
  from free-form queries}}.
\newblock In \bibinfo{editor}{Jonathan \surnamestart Aldrich\surnameend} \&
  \bibinfo{editor}{Patrick \surnamestart Eugster\surnameend}, editors: {\sl
  \bibinfo{booktitle}{OOPSLA}}, \bibinfo{publisher}{{ACM}}, pp.
  \bibinfo{pages}{416--432}, \doi{10.1145/2814270.2814295}.

\bibitemdeclare{inproceedings}{GveroETAL13CompleteCompletionUsingTypesWeights}
\bibitem{GveroETAL13CompleteCompletionUsingTypesWeights}
\bibinfo{author}{Tihomir \surnamestart Gvero\surnameend},
  \bibinfo{author}{Viktor \surnamestart Kuncak\surnameend},
  \bibinfo{author}{Ivan \surnamestart Kuraj\surnameend} \&
  \bibinfo{author}{Ruzica \surnamestart Piskac\surnameend}
  (\bibinfo{year}{2013}): \emph{\bibinfo{title}{Complete completion using types
  and weights}}.
\newblock In \bibinfo{editor}{Hans{-}Juergen \surnamestart Boehm\surnameend} \&
  \bibinfo{editor}{Cormac \surnamestart Flanagan\surnameend}, editors: {\sl
  \bibinfo{booktitle}{PLDI}}, \bibinfo{publisher}{{ACM}}, pp.
  \bibinfo{pages}{27--38}, \doi{10.1145/2462156.2462192}.

\bibitemdeclare{inproceedings}{DBLP:conf/cav/GveroKP11}
\bibitem{DBLP:conf/cav/GveroKP11}
\bibinfo{author}{Tihomir \surnamestart Gvero\surnameend},
  \bibinfo{author}{Viktor \surnamestart Kuncak\surnameend} \&
  \bibinfo{author}{Ruzica \surnamestart Piskac\surnameend}
  (\bibinfo{year}{2011}): \emph{\bibinfo{title}{Interactive Synthesis of Code
  Snippets}}.
\newblock In: {\sl \bibinfo{booktitle}{Computer Aided Verification - 23rd
  International Conference, {CAV} 2011, Snowbird, UT, USA, July 14-20, 2011.
  Proceedings}}, pp. \bibinfo{pages}{418--423},
  \doi{10.1007/978-3-642-22110-1_33}.

\bibitemdeclare{article}{DBLP:journals/corr/InalaQLS15}
\bibitem{DBLP:journals/corr/InalaQLS15}
\bibinfo{author}{Jeevana~Priya \surnamestart Inala\surnameend},
  \bibinfo{author}{Xiaokang \surnamestart Qiu\surnameend}, \bibinfo{author}{Ben
  \surnamestart Lerner\surnameend} \& \bibinfo{author}{Armando \surnamestart
  Solar{-}Lezama\surnameend} (\bibinfo{year}{2015}): \emph{\bibinfo{title}{Type
  Assisted Synthesis of Recursive Transformers on Algebraic Data Types}}.
\newblock {\sl \bibinfo{journal}{CoRR}} \bibinfo{volume}{abs/1507.05527}.
\newblock \urlprefix\url{http://arxiv.org/abs/1507.05527}.

\bibitemdeclare{phdthesis}{Kneuss-Thesis}
\bibitem{Kneuss-Thesis}
\bibinfo{author}{Etienne \surnamestart Kneuss\surnameend}
  (\bibinfo{year}{2016}): \emph{\bibinfo{title}{Deductive Synthesis and
  Repair}}.
\newblock Ph.D. thesis, \bibinfo{school}{EPFL}, \doi{10.5075/epfl-thesis-6878}.

\bibitemdeclare{inproceedings}{KneussKoukoutosKuncak15DeductiveProgramRepair}
\bibitem{KneussKoukoutosKuncak15DeductiveProgramRepair}
\bibinfo{author}{Etienne \surnamestart Kneuss\surnameend},
  \bibinfo{author}{Manos \surnamestart Koukoutos\surnameend} \&
  \bibinfo{author}{Viktor \surnamestart Kuncak\surnameend}
  (\bibinfo{year}{2015}): \emph{\bibinfo{title}{Deductive Program Repair}}.
\newblock In \bibinfo{editor}{Daniel \surnamestart Kroening\surnameend} \&
  \bibinfo{editor}{Corina~S. \surnamestart Pasareanu\surnameend}, editors: {\sl
  \bibinfo{booktitle}{CAV}}, {\sl \bibinfo{series}{LNCS}}
  \bibinfo{volume}{9207}, \bibinfo{publisher}{Springer}, pp.
  \bibinfo{pages}{217--233}, \doi{10.1007/978-3-319-21668-3_13}.

\bibitemdeclare{inproceedings}{KneussETAL13SynthesisModuloRecursiveFunctions}
\bibitem{KneussETAL13SynthesisModuloRecursiveFunctions}
\bibinfo{author}{Etienne \surnamestart Kneuss\surnameend},
  \bibinfo{author}{Ivan \surnamestart Kuraj\surnameend},
  \bibinfo{author}{Viktor \surnamestart Kuncak\surnameend} \&
  \bibinfo{author}{Philippe \surnamestart Suter\surnameend}
  (\bibinfo{year}{2013}): \emph{\bibinfo{title}{Synthesis modulo recursive
  functions}}.
\newblock In \bibinfo{editor}{Antony~L. \surnamestart Hosking\surnameend},
  \bibinfo{editor}{Patrick~Th. \surnamestart Eugster\surnameend} \&
  \bibinfo{editor}{Cristina~V. \surnamestart Lopes\surnameend}, editors: {\sl
  \bibinfo{booktitle}{OOPSLA}}, \bibinfo{publisher}{ACM}, pp.
  \bibinfo{pages}{407--426}, \doi{10.1145/2509136.2509555}.

\bibitemdeclare{article}{DBLP:journals/toplas/MannaW80}
\bibitem{DBLP:journals/toplas/MannaW80}
\bibinfo{author}{Zohar \surnamestart Manna\surnameend} \&
  \bibinfo{author}{Richard~J. \surnamestart Waldinger\surnameend}
  (\bibinfo{year}{1980}): \emph{\bibinfo{title}{A Deductive Approach to Program
  Synthesis}}.
\newblock {\sl \bibinfo{journal}{{ACM} Trans. Program. Lang. Syst.}}
  \bibinfo{volume}{2}(\bibinfo{number}{1}), pp. \bibinfo{pages}{90--121},
  \doi{10.1145/357084.357090}.

\bibitemdeclare{inproceedings}{MouraBjoerner08Z3EfficientSmtSolver}
\bibitem{MouraBjoerner08Z3EfficientSmtSolver}
\bibinfo{author}{Leonardo~Mendon{\c{c}}a \surnamestart de~Moura\surnameend} \&
  \bibinfo{author}{Nikolaj \surnamestart Bj{\o}rner\surnameend}
  (\bibinfo{year}{2008}): \emph{\bibinfo{title}{{Z3:} An Efficient {SMT}
  Solver}}.
\newblock In \bibinfo{editor}{C.~R. \surnamestart Ramakrishnan\surnameend} \&
  \bibinfo{editor}{Jakob \surnamestart Rehof\surnameend}, editors: {\sl
  \bibinfo{booktitle}{TACAS}}, {\sl \bibinfo{series}{LNCS}}
  \bibinfo{volume}{4963}, \bibinfo{publisher}{Springer}, pp.
  \bibinfo{pages}{337--340}, \doi{10.1007/978-3-540-78800-3_24}.

\bibitemdeclare{inproceedings}{PeiETAL15AutomatedProgramRepairIntegratedDevelopmentEnvironment}
\bibitem{PeiETAL15AutomatedProgramRepairIntegratedDevelopmentEnvironment}
\bibinfo{author}{Yu~\surnamestart Pei\surnameend}, \bibinfo{author}{Carlo~A.
  \surnamestart Furia\surnameend}, \bibinfo{author}{Mart{\'{\i}}n \surnamestart
  Nordio\surnameend} \& \bibinfo{author}{Bertrand \surnamestart
  Meyer\surnameend} (\bibinfo{year}{2015}): \emph{\bibinfo{title}{Automated
  Program Repair in an Integrated Development Environment}}.
\newblock In \bibinfo{editor}{Antonia \surnamestart Bertolino\surnameend},
  \bibinfo{editor}{Gerardo \surnamestart Canfora\surnameend} \&
  \bibinfo{editor}{Sebastian~G. \surnamestart Elbaum\surnameend}, editors: {\sl
  \bibinfo{booktitle}{ICSE}}, \bibinfo{publisher}{{IEEE} Computer Society}, pp.
  \bibinfo{pages}{681--684}, \doi{10.1109/ICSE.2015.222}.

\bibitemdeclare{inproceedings}{DBLP:conf/pldi/PerelmanGBG12}
\bibitem{DBLP:conf/pldi/PerelmanGBG12}
\bibinfo{author}{Daniel \surnamestart Perelman\surnameend},
  \bibinfo{author}{Sumit \surnamestart Gulwani\surnameend},
  \bibinfo{author}{Thomas \surnamestart Ball\surnameend} \&
  \bibinfo{author}{Dan \surnamestart Grossman\surnameend}
  (\bibinfo{year}{2012}): \emph{\bibinfo{title}{Type-directed completion of
  partial expressions}}.
\newblock In: {\sl \bibinfo{booktitle}{{ACM} {SIGPLAN} Conference on
  Programming Language Design and Implementation, {PLDI}}}, pp.
  \bibinfo{pages}{275--286}, \doi{10.1145/2254064.2254098}.

\bibitemdeclare{inproceedings}{PerelmanETAL14TestdrivenSynthesis}
\bibitem{PerelmanETAL14TestdrivenSynthesis}
\bibinfo{author}{Daniel \surnamestart Perelman\surnameend},
  \bibinfo{author}{Sumit \surnamestart Gulwani\surnameend},
  \bibinfo{author}{Dan \surnamestart Grossman\surnameend} \&
  \bibinfo{author}{Peter \surnamestart Provost\surnameend}
  (\bibinfo{year}{2014}): \emph{\bibinfo{title}{Test-driven synthesis}}.
\newblock In \bibinfo{editor}{Michael F.~P. \surnamestart O'Boyle\surnameend}
  \& \bibinfo{editor}{Keshav \surnamestart Pingali\surnameend}, editors: {\sl
  \bibinfo{booktitle}{PLDI}}, \bibinfo{publisher}{{ACM}},
  p.~\bibinfo{pages}{43}, \doi{10.1145/2594291.2594297}.

\bibitemdeclare{inproceedings}{DBLP:conf/pldi/PolikarpovaKS16}
\bibitem{DBLP:conf/pldi/PolikarpovaKS16}
\bibinfo{author}{Nadia \surnamestart Polikarpova\surnameend},
  \bibinfo{author}{Ivan \surnamestart Kuraj\surnameend} \&
  \bibinfo{author}{Armando \surnamestart Solar{-}Lezama\surnameend}
  (\bibinfo{year}{2016}): \emph{\bibinfo{title}{Program synthesis from
  polymorphic refinement types}}.
\newblock In: {\sl \bibinfo{booktitle}{Proceedings of the 37th {ACM} {SIGPLAN}
  Conference on Programming Language Design and Implementation, {PLDI} 2016,
  Santa Barbara, CA, USA, June 13-17, 2016}}, pp. \bibinfo{pages}{522--538},
  \doi{10.1145/2908080.2908093}.

\bibitemdeclare{inproceedings}{PolozovGulwani15FlashmetaFrameworkInductiveProgramSynthesis}
\bibitem{PolozovGulwani15FlashmetaFrameworkInductiveProgramSynthesis}
\bibinfo{author}{Oleksandr \surnamestart Polozov\surnameend} \&
  \bibinfo{author}{Sumit \surnamestart Gulwani\surnameend}
  (\bibinfo{year}{2015}): \emph{\bibinfo{title}{FlashMeta: a framework for
  inductive program synthesis}}.
\newblock In \bibinfo{editor}{Jonathan \surnamestart Aldrich\surnameend} \&
  \bibinfo{editor}{Patrick \surnamestart Eugster\surnameend}, editors: {\sl
  \bibinfo{booktitle}{OOPSLA}}, \bibinfo{publisher}{{ACM}}, pp.
  \bibinfo{pages}{107--126}, \doi{10.1145/2814270.2814310}.

\bibitemdeclare{inproceedings}{ReinkingPiskac15TypedirectedApproachToProgramRepair}
\bibitem{ReinkingPiskac15TypedirectedApproachToProgramRepair}
\bibinfo{author}{Alex \surnamestart Reinking\surnameend} \&
  \bibinfo{author}{Ruzica \surnamestart Piskac\surnameend}
  (\bibinfo{year}{2015}): \emph{\bibinfo{title}{A Type-Directed Approach to
  Program Repair}}.
\newblock In \bibinfo{editor}{Daniel \surnamestart Kroening\surnameend} \&
  \bibinfo{editor}{Corina~S. \surnamestart Pasareanu\surnameend}, editors: {\sl
  \bibinfo{booktitle}{CAV}}, {\sl \bibinfo{series}{LNCS}}
  \bibinfo{volume}{9206}, \bibinfo{publisher}{Springer}, pp.
  \bibinfo{pages}{511--517}, \doi{10.1007/978-3-319-21690-4_35}.

\bibitemdeclare{inproceedings}{liquid}
\bibitem{liquid}
\bibinfo{author}{Patrick~M. \surnamestart Rondon\surnameend},
  \bibinfo{author}{Ming \surnamestart Kawaguci\surnameend} \&
  \bibinfo{author}{Ranjit \surnamestart Jhala\surnameend}
  (\bibinfo{year}{2008}): \emph{\bibinfo{title}{Liquid Types}}.
\newblock In: {\sl \bibinfo{booktitle}{Proceedings of the 29th ACM SIGPLAN
  Conference on Programming Language Design and Implementation}},
  \bibinfo{series}{PLDI '08}, \bibinfo{publisher}{ACM}, \bibinfo{address}{New
  York, NY, USA}, pp. \bibinfo{pages}{159--169}, \doi{10.1145/1375581.1375602}.

\bibitemdeclare{inproceedings}{sketch}
\bibitem{sketch}
\bibinfo{author}{Armando \surnamestart Solar-Lezama\surnameend},
  \bibinfo{author}{Liviu \surnamestart Tancau\surnameend},
  \bibinfo{author}{Rastislav \surnamestart Bodik\surnameend},
  \bibinfo{author}{Sanjit \surnamestart Seshia\surnameend} \&
  \bibinfo{author}{Vijay \surnamestart Saraswat\surnameend}
  (\bibinfo{year}{2006}): \emph{\bibinfo{title}{Combinatorial Sketching for
  Finite Programs}}.
\newblock In: {\sl \bibinfo{booktitle}{Proceedings of the 12th International
  Conference on Architectural Support for Programming Languages and Operating
  Systems}}, \bibinfo{series}{ASPLOS XII}, \bibinfo{publisher}{ACM},
  \bibinfo{address}{New York, NY, USA}, pp. \bibinfo{pages}{404--415},
  \doi{10.1145/1168857.1168907}.

\bibitemdeclare{phdthesis}{suter-thesis}
\bibitem{suter-thesis}
\bibinfo{author}{Philippe \surnamestart Suter\surnameend}
  (\bibinfo{year}{2012}): \emph{\bibinfo{title}{Programming with
  Specifications}}.
\newblock Ph.D. thesis, \bibinfo{school}{EPFL}, \doi{10.5075/epfl-thesis-5581}.

\bibitemdeclare{inproceedings}{SuterKoeksalKuncak11SatisfiabilityModuloRecursivePrograms}
\bibitem{SuterKoeksalKuncak11SatisfiabilityModuloRecursivePrograms}
\bibinfo{author}{Philippe \surnamestart Suter\surnameend},
  \bibinfo{author}{Ali~Sinan \surnamestart K{\"o}ksal\surnameend} \&
  \bibinfo{author}{Viktor \surnamestart Kuncak\surnameend}
  (\bibinfo{year}{2011}): \emph{\bibinfo{title}{Satisfiability Modulo Recursive
  Programs}}.
\newblock In \bibinfo{editor}{Eran \surnamestart Yahav\surnameend}, editor:
  {\sl \bibinfo{booktitle}{SAS}}, {\sl \bibinfo{series}{LNCS}}
  \bibinfo{volume}{6887}, \bibinfo{publisher}{Springer}, pp.
  \bibinfo{pages}{298--315}, \doi{10.1007/978-3-642-23702-7_23}.

\bibitemdeclare{inproceedings}{DBLP:conf/pldi/UdupaRDMMA13}
\bibitem{DBLP:conf/pldi/UdupaRDMMA13}
\bibinfo{author}{Abhishek \surnamestart Udupa\surnameend},
  \bibinfo{author}{Arun \surnamestart Raghavan\surnameend},
  \bibinfo{author}{Jyotirmoy~V. \surnamestart Deshmukh\surnameend},
  \bibinfo{author}{Sela \surnamestart Mador{-}Haim\surnameend},
  \bibinfo{author}{Milo M.~K. \surnamestart Martin\surnameend} \&
  \bibinfo{author}{Rajeev \surnamestart Alur\surnameend}
  (\bibinfo{year}{2013}): \emph{\bibinfo{title}{{TRANSIT:} specifying protocols
  with concolic snippets}}.
\newblock In: {\sl \bibinfo{booktitle}{{ACM} {SIGPLAN} Conference on
  Programming Language Design and Implementation, {PLDI}}}, pp.
  \bibinfo{pages}{287--296}, \doi{10.1145/2462156.2462174}.

\bibitemdeclare{inproceedings}{wei2010automated}
\bibitem{wei2010automated}
\bibinfo{author}{Yi~\surnamestart Wei\surnameend},
  \bibinfo{author}{Yu~\surnamestart Pei\surnameend}, \bibinfo{author}{Carlo~A
  \surnamestart Furia\surnameend}, \bibinfo{author}{Lucas~S \surnamestart
  Silva\surnameend}, \bibinfo{author}{Stefan \surnamestart
  Buchholz\surnameend}, \bibinfo{author}{Bertrand \surnamestart
  Meyer\surnameend} \& \bibinfo{author}{Andreas \surnamestart
  Zeller\surnameend} (\bibinfo{year}{2010}): \emph{\bibinfo{title}{Automated
  fixing of programs with contracts}}.
\newblock In: {\sl \bibinfo{booktitle}{Proceedings of the 19th international
  symposium on Software testing and analysis}}, \bibinfo{organization}{ACM},
  pp. \bibinfo{pages}{61--72}, \doi{10.1145/1831708.1831716}.

\end{thebibliography}
